\documentclass[aps,prd,preprint,groupedaddress]{revtex4-1}
\usepackage{extraipa}
\usepackage{amsmath}
\usepackage{amsfonts}
\usepackage{amssymb}
\usepackage{bbold}

\newcommand{\etal}{\textit{et al}.}

\begin{document}

\title{Note on stability and microcausality in Lorentz violating antisymmetric tensor}

\author{Sandeep Aashish}%
\email[]{sandeepa16@iiserb.ac.in}


\affiliation{Department of Physics, Indian Institute of Science Education and Research, Bhopal 462066, India}

%


\date{\today}

\begin{abstract}
The fundamental issues of microcausality and energy positivity conditions are important to be investigated in the context of spontaneous Lorentz violating theories. We check the microcausality and energy positivity conditions for a free rank-2 antisymmetric tensor field with spontaneous Lorentz violating term and a classically equivalent vector theory from dispersion relations and propagators, and find that the two theories lead to different conditions. While the antisymmetric tensor theory satisfies energy positivity condition and can satisfy microcausality condition for a particular choice of vacuum value, the vector theory violates both energy positivity and microcausality conditions.
\end{abstract}



\maketitle

\section{Introduction}
Causality, and stability (or energy positivity), are some of the most fundamental requirements for a physical theory \cite{schroer2011}. In classical electrodynamics for example, the statement of causality translates to fixing the time-order of cause and effect over classical distances, wherein the effect always occurs after the cause, and is often set by hand through the use of retarded green's functions \cite{jackson1998}. In quantum field theory, it is possible to define a more robust statement of causality valid over microscopic scales, by demanding that any two observables separated by a space-like distance must commute thereby ensuring that measurements over spacelike distances do not affect each other and thus cannot be causally related. This statement in quantum field theory is known as microcausality \cite{peskin1995}. Stability, strictly in the context of energy positivity condition, in a quantum or classical theory requires that a positive energy solution must stay positive, and is determined from the dispersion relations of a given theory \cite{zwicky2016}.

These fundamental requirements, especially microcausality, is satisfied by any Lorentz invariant quantum field theory by construction since it turns out that the integrals appearing in the commutation relation of fields vanish identically due to Lorentz invariance property \cite{peskin1995}. However, it becomes important to check microcausality and stability in case of a Lorentz non-invariant theory and have been part of several past studies \cite{kostelecky2001b,dubovsky2008,bros2002,rashidi2007}. Similarly, in theories with spontaneous Lorentz violation, the issue of microcausality is more subtle and often introduces constraints on the vacuum values of field. For instance, such studies were performed for vector bumblebee models in Refs. \cite{maluf2014,bluhm2008b}. For a dirac spinor with spontaneous Lorentz violation, which is part of SME, the conditions on the LV coefficients for microcausality and stability were analyzed by Kostelecky \etal \cite{kostelecky2001b}. 

While several indirect results for antisymmetric tensor fields with spontaneous Lorentz violation do exist (see, for example Refs. \cite{altschul2010,seifert2010a,seifert2010b,seifert2019}) in different cosmological and formal contexts, the purpose here is to bridge the gap in existing literature, by addressing the stability and microcausality issues using dispersion relation and the propagators in Minkowski spacetime. In addition to microcausality, the issue of stability is especially relevant in case of antisymmetric tensor field because of recent results by Seifert \cite{seifert2019} showing that field fluctuations about the vacuum manifold may not be stable. We also consider these issues for an equivalent (in flat spacetime) theory of a vector field. 

The organization is as follows. In Sec. \ref{sec2}, we lay out the background results for the model of antisymmetric tensor field under consideration. The dispersion relations and the propagators to address the question of microcausality in Sec. \ref{sec3}. We conclude in Sec. \ref{sec5}

\section{\label{sec2}Background}
In theories without any Lorentz violation, Lorentz symmetry ensures that upon either particle or coordinate (or observer) Lorentz transformations or both, the physics does not change. So, there are essentially two ways one can break Lorentz symmetry in a theory, either explicitly in which case both particle and frame transformations break this symmetry, or spontaneously wherein only the particle transformations break the symmetry. This also ensures that the fundamental requirement that a physical theory must be independent of observers is intact. Moreover, due to problems with Bianchi identities in case of explicit Lorentz violation \cite{altschul2010}, spontaneous Lorentz violation is a preferable way to study standard model extension (SME) theories \cite{kostelecky1995,colladay1997,colladay1998,kostelecky2004,bluhm2006}, while also having potentially observable experimental signatures. 

We consider a model of rank-2 antisymmetric tensor field $B_{\mu\nu}$ that incorporates spontaneous Lorentz violation via a nonzero vacuum expectation value of the field, 
\begin{eqnarray}
    \label{vev}
    \langle B_{\mu\nu}\rangle = b_{\mu\nu}. 
\end{eqnarray}
Choosing, or fixing, the vacuum value is equivalent to breaking the particle Lorentz symmetry. Since observer Lorentz symmetry still holds, one can choose to work in a special frame where 
the vacuum value $b_{\mu\nu}$ takes a particular form given by \cite{altschul2010}
\begin{eqnarray}
    \label{slva1}
    b_{\mu\nu} = 
    \left(\begin{matrix}
    0 & -a & 0 & 0\\
    a & 0 & 0 & 0\\
    0 & 0 & 0 & b\\
    0 & 0 & -b & 0
    \end{matrix}\right),
\end{eqnarray}
where $a$ and $b$ are real numbers satisfying the requirement that at least one of the quantities $X_{1}=-2(a^2-b^2)$ and $X_{2}=4ab$ is nonzero. Although we work with the general  $b_{\mu\nu}$ tensor for the most part, the structure in Eq. (\ref{slva1}) determines stability and microcausality conditions. Moreover, as we will note later on, it turns out that one must choose $a=0$ for a spherically symmetric nontrivial solution of the equation of motion of $B_{\mu\nu}$ with the potential considered here, to approach vacuum value asymptotically \cite{seifert2010a,seifert2010b}. 

The choice of potential is largely driven by the requirement of constructing the simplest derivative-free Lorentz invariant term involving the vacuum value $b_{\mu\nu}$, and is given by,
\begin{eqnarray}
    \label{slva2}
    V(B) = \frac{1}{16}m^2 \Big(B_{\mu\nu}B^{\mu\nu} - b_{\mu\nu}b^{\mu\nu}\Big)^{2},
\end{eqnarray}
which has terms upto quartic order in fields. However, for the present purposes we restrict ourselves to only quadratic terms in $B_{\mu\nu}$, neglecting quartic and cubic fluctuations. For this reason, we instead work with small field fluctuations about the vacuum value $b_{\mu\nu}$, written as,
\begin{eqnarray}
    \label{slva3}
    B_{\mu\nu} = b_{\mu\nu} + \tilde{B}_{\mu\nu}
\end{eqnarray}
where we assume $|| \tilde{B}_{\mu\nu}|| \ll ||b_{\mu\nu}||$, so that the higher order contributions can be ignored. In terms of these fluctuations, the Lagrangian upto quadratic order in field fluctuations is given by,
\begin{eqnarray}
    \label{lag}
      \mathcal{L} = -\frac{1}{12}H_{\mu\nu\lambda}H^{\mu\nu\lambda} - \dfrac{1}{4}m^{2} \Big(b_{\mu\nu}{B}^{\mu\nu}\Big)^{2},
\end{eqnarray} 
where for notational convenience we write $B_{\mu\nu}$ in place of $\tilde{B}_{\mu\nu}$ here and throughout. The first term is the gauge invariant kinetic term, defined as, $H_{\mu\nu\lambda} = \partial_{\mu}B_{\nu\lambda} + \partial_{\lambda}B_{\mu\nu} + \partial_{\nu}B_{\lambda\mu}$, and is invariant under the transformations $B^{\mu\nu} \longrightarrow B^{\mu\nu} + \nabla_{\mu}\xi_{\nu} - \nabla_{\nu}\xi_{\mu}$ where $\xi_{\mu}$ is the gauge parameter. 

Naively, one would treat the Lagrangian (\ref{lag}) as a non-gauge invariant one while quantizing, and deriving the propagators. While it does not affect our results in the present case, we note that the gauge-invariance of the kinetic term in Eq. (\ref{lag}) does introduce redundancies in the energy spectrum apparent in the Hamiltonian description \cite{buchbinder2007}. Hence, we opt for a more consistent Lagrangian through the application of St{\"u}ckelberg procedure \cite{stuckelberg1957}, where the overall gauge symmetry is restored in the Lagrangian via the introduction of a so-called St{\"u}ckelberg field. This does not change the theory, or the degrees of freedom (see \cite{buchbinder1992,buchbinder2008,aashish2018a}, for example), rather the original theory is recovered in a special gauge where the St{\"u}ckelberg field is set to zero. The quantization procedure then involves the usual Faddeev-Popov method for all fields including St{\"u}ckelberg field(s) for fixing the gauge. We skip the calculational details carried out elsewhere \cite{aashish2019b,aashish2018b}, and present the final gauge-fixed Lagrangian for the original theory (\ref{lag}), 
\begin{eqnarray}
    \label{slvb3}
\mathcal{L}_{GF} = -\dfrac{1}{12}H_{\mu\nu\lambda}H^{\mu\nu\lambda} - \frac{1}{4}m^{2}\Big(b_{\mu\nu}B^{\mu\nu}\Big)^{2} - \dfrac{1}{4}\Big(b_{\mu\nu}F^{\mu\nu}\Big)^{2} - \frac{1}{2}\Big(b_{\mu\nu}b_{\rho\sigma}\nabla^{\mu}B^{\rho\sigma}\Big)^{2} \nonumber \\ - \frac{1}{2}m^{2}C_{\nu}C^{\nu} - \frac{1}{2}(\nabla_{\mu}\Phi)^{2} - \frac{1}{2}(\nabla^{\mu}C_{\mu})^{2} - \frac{1}{2}m^{2}\Phi^{2}.
\end{eqnarray}
where $C_{\mu}$ and $\Phi$ are the new fields introduced due to the St{\"u}ckelberg method. The presence of scalar field $\Phi$ in Eq. (\ref{slvb3}) is a direct consequence of gauge-fixing of $C_{\mu}$, and explicitly displays a scalar degree of freedom that remains hidden in the original Lagrangian (\ref{lag}). Note that, the last term in the first line of Eq. (\ref{slvb3}) is absent from antisymmetric tensor Lagrangian if one does not treat Eq. (\ref{lag}) with the St{\"u}ckelberg procedure. Though, the result for propagator of $B_{\mu\nu}$ is unaffected since the additional term does not contribute \cite{maluf2019,aashish2019b}.

\section{\label{sec3}Dispersion relations and microcausality}
The equation of motion for the free antisymmetric tensor field in theory Eq. (\ref{slvb3}) is given by:
\begin{eqnarray}
    \label{eom}
    \mathcal{O}^{X}_{\mu\nu,\alpha\beta}B^{\alpha\beta} = 0, 
\end{eqnarray}
where 
\begin{eqnarray}
\label{apr1}
\mathcal{O}^{X}_{\mu\nu,\alpha\beta} &=& \frac{\Box}{2}(\eta_{\mu\alpha}\eta_{\nu\beta}-\eta_{\mu\beta}\eta_{\nu\alpha})  + \frac{1}{2}(\partial_{\mu}\partial_{\beta}\eta_{\nu\alpha} - \partial_{\nu}\partial_{\beta}\eta_{\mu\alpha} - \partial_{\mu}\partial_{\alpha}\eta_{\nu\beta} + \partial_{\nu}\partial_{\alpha}\eta_{\mu\beta})\nonumber\\ && -\,\big(m^2 + 2 (b_{\rho\sigma}\partial^{\rho})^2\big) b_{\mu\nu}b_{\alpha\beta}.
\end{eqnarray}
The dispersion relation is derived from the equation of motion by going to the Fourier space so that Eq. (\ref{eom}) is rewritten in terms of the frequency (or energy) modes and momenta, solving which after application of appropriate constraints leads to the dispersion relation. Substituting $B_{\alpha\beta} = \int d^{4}k \tilde{B}_{\alpha\beta} e^{-ik\cdot x}$ in Eq. (\ref{eom}), we get the momentum space representation of $\mathcal{O}^{X}_{\mu\nu,\alpha\beta}$,
\begin{eqnarray}
    \label{opmom}
    \mathcal{O}_{\mu\nu,\alpha\beta} &=& -\frac{k^2}{2}(\eta_{\mu\alpha}\eta_{\nu\beta}-\eta_{\mu\beta}\eta_{\nu\alpha})  + \frac{1}{2}(-k_{\mu}k_{\beta}\eta_{\nu\alpha} + k_{\nu}k_{\beta}\eta_{\mu\alpha} + k_{\mu}k_{\alpha}\eta_{\nu\beta} - k_{\nu}k_{\alpha}\eta_{\mu\beta})\nonumber\\ && - \big(m^2 - 2 (b_{\rho\sigma}k^{\rho})^2\big) b_{\mu\nu}b_{\alpha\beta}.
\end{eqnarray}
In general, an operator equation of the form $\mathcal{O}\psi=0$ in momentum space gives rise to the dispersion relation upon solving $\det(\mathcal{O})=0$ for the constraints. Note however that, in case of Eq. (\ref{eom}), the operator $\mathcal{O}_{\mu\nu\alpha\beta}$ is rank-4. As a result the dispersion relation derivation requires some manipulations. We use Maple \cite{maple} for these manipulations. Eq. (\ref{eom}) essentially is a set of multiple equations corresponding to each choice of ($\mu,\nu$). Now, for a given choice of ($\mu=i,\nu=j$) with $i=1..4, j=1..4$, we then have a rank-2 (antisymmetric) operator $\mathcal{O}_{ij,\alpha\beta}$ in momentum space acting on the field $B^{\alpha\beta}$. For simplicity, we choose the momentum components such that $k^{\mu}=(k^0,0,0,k^{3})$ without loss of generality. Hence, for each equation the constraint condition is given by $\det(\mathcal{O}_{ij,\alpha\beta})=0$.
Out of a total of five distinct equations (instead of six, due to the choice of $k^{\mu}$; however it does not affect the final constraint condition), it turns out that, 
\begin{eqnarray}
    \label{det0}
    \det(\mathcal{O}_{13,\alpha\beta}) = \det(\mathcal{O}_{23,\alpha\beta}) = \det(\mathcal{O}_{24,\alpha\beta}) = 0. 
\end{eqnarray}
The remaining non-zero determinants are as follows:
\begin{eqnarray}
    \label{nzdet}
    \det(\mathcal{O}_{12,\alpha\beta}) &=& -1/2\, \left( -2\,{a}^{4}{{\it k0}}^{2}-2\,{a}^{2}{b}^{2}{{\it k3}}^{2
}+{a}^{2}{m}^{2}-1/2\,{{\it k3}}^{2} \right) \nonumber \\ && \times  \left( 4\,{a}^{4}{{\it 
k0}}^{2}+4\,{a}^{2}{b}^{2}{{\it k3}}^{2}-2\,{a}^{2}{m}^{2}+{{\it k3}}^
{2} \right) \nonumber \\ && \times \left( 2\,{a}^{2}{{\it k0}}^{2}+2\,{b}^{2}{{\it k3}}^{2}-
{m}^{2} \right) ^{2}{a}^{2}{b}^{2};\\
\det(\mathcal{O}_{34,\alpha\beta}) &=& -1/4\, \left( -2\,{a}^{2}{{\it k0}}^{2}-2\,{b}^{2}{{\it k3}}^{2}+{m}^{
2} \right) {a}^{2}{b}^{2} \nonumber \\ && \times \left( 2\,{a}^{2}{{\it k0}}^{2}+2\,{b}^{2}{{
\it k3}}^{2}-{m}^{2} \right) \nonumber \\ && \times \left( 4\,{a}^{2}{b}^{2}{{\it k0}}^{2}+4
\,{b}^{4}{{\it k3}}^{2}-2\,{b}^{2}{m}^{2}+{{\it k0}}^{2} \right) ^{2};
\end{eqnarray}
We solve all these equations for the zeroth component of momenta, $k^{0}$, to obtain the constraint equation. We obtain, 
\begin{eqnarray}
    \label{vval}
    \big(k^{0}\big)^{2} = \dfrac{-4 \big(k^{3}\big)^2 b^2 + 2 m^2}{4 a^2}. 
\end{eqnarray}
Using the tensor structure of the vacuum value $b_{\mu\nu}$ in Eq. (\ref{vval}), we rewrite the above equation as,
\begin{eqnarray}
    \label{constraint1}
    (b_{\mu\nu}k^{\nu})^2 = -\dfrac{m^2}{2}. 
\end{eqnarray}
Substituting the constraint Eq. (\ref{constraint1}) in Eq. (\ref{eom}), the last term in the operator $\mathcal{O}_{\mu\nu\alpha\beta}$ vanishes. As a result, we get,
\begin{eqnarray}
    \label{disp0}
    \Big[-\dfrac{k^2}{2}(\eta_{\alpha\mu}\eta_{\beta\nu}-\eta_{\beta\mu}\eta_{\alpha\nu}) + \frac{1}{2}k_{\nu}k_{\beta}\eta_{\alpha\mu} - \frac{1}{2}k_{\mu}k_{\beta}\eta_{\alpha\nu} - \frac{1}{2}k_{\nu}k_{\alpha}\eta_{\beta\mu} + \frac{1}{2}k_{\mu}k_{\alpha}\eta_{\beta\nu}\Big]B^{\alpha\beta} = 0.
\end{eqnarray}
Here, $k^2$ represents the norm of four momentum $k^{\mu}$, while $\eta_{\mu\nu}$ is the Minkowski metric. A common trick while deriving a dispersion relation is to try and eliminate the free indices so that one ends up with a scalar equation with separable momenta and field terms. With a similar intention, we multiply Eq. (\ref{disp0}) with $b_{\rho\mu}b^{\rho}_{\nu}$. This results in, 
\begin{eqnarray}
    \label{disp1}
    -k^2 b_{\rho[\alpha}b^{\rho}_{\beta]}B^{\alpha\beta} + b_{\rho\nu}k^{\nu}b^{\rho}_{[\alpha}k_{\beta]}B^{\alpha\beta} - b_{\rho\mu}k^{\mu}b^{\rho}_{[\alpha}k_{\beta]}B^{\alpha\beta} = 0,
\end{eqnarray}
where the brackets "[" indicate antisymmetrization. 
Clearly, the last two terms in Eq. (\ref{disp1}) cancel each other out to yield the dispersion relation, 
\begin{eqnarray}
    \label{disp2}
    k^2 = 0. 
\end{eqnarray}
It is straightforward to infer that there is no energy positivity issue here, and it describes a massless propagating mode. 

Next, we investigate the dispersion relations for a classically equivalent vector theory. The classical and quantum equivalence of 2-form with a vector field was studied in Ref. \cite{aashish2018b,aashish2019b}. The Lagrangian is given by, 
\begin{eqnarray}
    \label{eq3}
    \mathcal{L} = \dfrac{1}{4}\left(\tilde{b}_{\mu\nu}F^{\mu\nu}\right)^{2} - \frac{1}{2} m^{2} A^{\mu}A_{\mu},
    \end{eqnarray}
where $\tilde{b}_{\mu\nu} = \dfrac{1}{2}\epsilon_{\mu\nu\rho\sigma}b^{\rho\sigma}$. 
We note that the theories are actually not quantum equivalent in a curved spacetime, but can be treated as equivalent theories in the present context since we work in Minkowski spacetime.

The equation of motion for the vector theory (\ref{eq3}) is given by, 
\begin{eqnarray}
    \label{veom}
    \mathcal{O}^{v}_{\mu\nu} A^{\nu} &=& 0. 
\end{eqnarray}
where the momentum space representation of $\mathcal{O}^{v}_{{\mu\nu}}$ is, 
\begin{eqnarray}
    \mathcal{O}^{v}_{{\mu\nu}} = \dfrac{1}{2}\big(\tilde{b}_{\sigma\mu}\tilde{b}_{\rho\nu}k^{\sigma}k^{\rho} + \tilde{b}_{\sigma\nu}\tilde{b}_{\rho\mu}k^{\sigma}k^{\rho}\big) + \dfrac{1}{2}k_{\mu}k_{\nu} - \dfrac{1}{2} m^{2}\eta_{\mu\nu}.
\end{eqnarray}
Since the operator in Eq. (\ref{veom}) is rank-2, it is straightforward to derive dispersion relations here. Using the same trick as above, we first multiply $k^{\mu}$ on both sides of Eq. (\ref{veom}), which yields,
\begin{eqnarray}
    \big[\frac{1}{2}\tilde{b}_{\nu\sigma}\tilde{b}_{\mu\rho}k^{\sigma}k^{\rho}k^{\mu} + \frac{1}{2}\tilde{b}_{\mu\sigma}\tilde{b}_{\nu\rho}k^{\sigma}k^{\rho}k^{\mu} - \frac{1}{2}(k^2 + m^2)k_{\nu}\big] A^{\nu} = 0.
\end{eqnarray}
Here the first two terms vanish identically because of the antisymmetric $\tilde{b}_{\mu\nu}$, so that, 
\begin{eqnarray}
    \label{vdisprel1}
    k^2 + m^2 = 0. 
\end{eqnarray}
Before analyzing the dispersion relation in Eq. (\ref{vdisprel1}), we first look at another possibility. Multiplying Eq. (\ref{veom}) with $\tilde{b}^{\mu}_{\ \alpha}k^{\alpha}$ on both sides, we get
\begin{eqnarray}
    \label{vdisprel2}
    \tilde{b}_{\nu\sigma}\tilde{b}_{\mu\rho}k^{\rho}k^{\sigma}\tilde{b}^{\mu}_{\alpha}k^{\alpha}A^{\nu} - \dfrac{1}{2}k_{\mu}\tilde{b}^{\mu}_{\alpha}k^{\alpha}k_{\nu}A^{\nu} - \dfrac{1}{2}m^2\eta_{\mu\nu}\tilde{b}^{\mu}_{\ \alpha}k^{\alpha}A^{\nu} = 0.
\end{eqnarray}
Once again, due to symmetry properties the second term in Eq. (\ref{vdisprel2}) vanishes, and separating out the scalar $\tilde{b}_{\nu \alpha}k^{\alpha}A^{\nu}$, we get another dispersion relation
\begin{eqnarray}
    \label{vdisp3}
    (\tilde{b}\cdot k)^{2} - \dfrac{1}{2}m^2 = 0. 
\end{eqnarray}
A few comments are in order about the dispersion relations (\ref{vdisprel1},\ref{vdisp3}). In Eq. (\ref{vdisprel1}) the negative sign of $m^2$ implies that $(k^0)^2 = \vec{k}^2 - m^2$ can be negative, thus violating energy positivity conditions. In contrast to the 2-form case Eq. (\ref{disp2}), the dispersion relations of classically equivalent vector theory describe massive modes. Moreover, Eq. (\ref{vdisprel1}) also describes a massive mode corresponding to the Lorentz violating mode which is absent in case of 2-forms. 
Infact, massive Lorentz violating modes do appear in 2-form case but as a constraint equation. 

The dispersion relations contain the information about energy positivity, but in order to analyze the propagating modes and eventually the microcausality condition, one must look at the propagators for concerned theories. Similar analyses have been undertaken for vector bumblebee models and also for the antisymmetric tensor \cite{kostelecky2001b,maluf2014,bluhm2008b}. The propagators for the model(s) under consideration here have already been calculated in Refs. \cite{aashish2019b,maluf2019}. Here we use these results to analyze the poles and check microcausality conditions. 

Using the expressions for the projection operators, the final explicit form of propagator for the 2-form field is\begin{eqnarray}
    \label{prop1}
    D_{\mu\nu\rho\sigma} &=& \dfrac{-\eta_{\nu\sigma}k_{\mu}k_{\rho} + \eta_{\mu\rho}k_{\nu}k_{\sigma} + \eta_{\nu\rho}k_{\mu}k_{\sigma} - \eta_{\mu\rho}k_{\nu}k_{\sigma}}{2k^4} + \dfrac{\eta_{\mu\rho}\eta_{\nu\sigma} - \eta_{\mu\sigma}\eta_{\nu\rho}}{2k^2} \nonumber \\ && + \dfrac{b_{\nu\alpha}b_{\rho\sigma}k^{\alpha}k_{\mu} - b_{\mu\alpha}b_{\rho\sigma}k^{\alpha}k_{\nu} + b_{\mu\nu}b_{\sigma\alpha}k^{\alpha}k_{\rho} - b_{\mu\nu}b_{\rho\alpha}k^{\alpha}k_{\sigma}}{2 k^2 (b\cdot k)^2};
\end{eqnarray}
and the propagator for the equivalent vector theory is,
\begin{eqnarray}
    \label{prop2}
    D^{v}_{\mu\nu} &=& -\dfrac{\eta_{\mu\nu}}{m^2} + \dfrac{k_{\mu}k_{\nu}}{m^2 k^2} - \dfrac{k_{\mu}k_{\nu}}{k^2 (k^2 + m^2)} + \dfrac{\tilde{b}_{\mu\alpha}b_{\nu\beta}k^{\alpha}k^{\beta}}{m^2 ((\tilde{b}\cdot k)^2 - \frac{m^2}{2})}. 
\end{eqnarray}

As expected from the dispersion relations, only the massless mode propagates in a 2-form theory with spontaneous Lorentz violation. However, there exists a non-physical pole $(b\cdot k)^2 = 0$ involving vacuum value $b_{\mu\nu}$ that does not appear in the dispersion relations. In terms of the momentum components, this non physical pole takes the form,
\begin{eqnarray}
    \label{antipole}
    a^2 ((k^0)^2 - (k^1)^2) + b^2 ((k^2)^2 + (k^3)^2) = 0. 
\end{eqnarray}
Clearly, the energy positivity condition is violated since $(k^{0})^{2}$ can be negative. However an interesting case is when one considers a vacuum structure inspired from the requirements for the existence of a monopole solution \cite{seifert2010a,seifert2010b}, i.e. putting $a=0$. Then, all energy terms disappear from the poles, and as a result all Lorentz violating modes have no meaningful contribution to the propagator. 

For the vector theory, similarly we have a massless pole $k^2=0$ in addition to the expected poles from dispersion relations, 
\begin{eqnarray}
    \label{vdipr1}
    k^2 = -m^2; \\
    \label{vdipr2}
    (\tilde{b}\cdot k)^2 = \dfrac{m^2}{2}.
\end{eqnarray}
Both poles in the vector propagator can violate energy positivity condition. For the Lorentz violating mode, we have
\begin{eqnarray}
    (k^0)^2 = -(k^1)^2 + \dfrac{a^2}{b^2}((k^2)^2 - (k^3)^2) - \dfrac{m^2}{2 b^2}. 
\end{eqnarray}
For the same case $a=0$, $(k^0)^2 < 0$ and thus violates energy positivity conditions. 
Notice the contrast in the dispersion relations (\ref{vdipr1},\ref{vdipr2}) and (\ref{disp2}) corresponding to supposedly equivalent tensor and vector theories respectively. The theories are have been shown to be equivalent in flat spacetime with respect to their effective actions \cite{aashish2018b}, and later shown to break the equivalence in curved spacetime \cite{aashish2019b}. However, in the present analysis which is also in flat spacetime, the theories are not equivalent at the level of dispersion relations.

To address microcausality, one needs to evaluate the commutator of two spacelike separated observables. In the present case, the fields $B_{\mu\nu}$ and $A_{\mu}$ are not observables themselves. Observables can be constructed out of gauge-invariant constructions of $B_{\mu\nu}$ and $A_{\mu}$, usually in combination with derivative operators $\partial_{\mu}$. However, the observables always commute and thus microcausality condition is always satisfied if the commutator of fields at spacelike separation vanishes. For any field $\mathcal{F}(x)$, the microcausality condition requires,
\begin{eqnarray}
    \label{com1}
    S(x-x') \equiv [\mathcal{F}(x),\mathcal{F}(x')] = 0; \quad (x-x')^{2} < 0. 
\end{eqnarray}
The commutator $S(x-x')$ can be written in terms of the retarded and advanced Green's functions \cite{peskin1995,kostelecky2001b}, and consequently can be expressed in terms of the integral 
\begin{eqnarray}
    S(z) = \int_{C} \dfrac{d^4 k}{(2\pi)^4} e^{-i k\cdot z} D(k),
\end{eqnarray}
where the contour $C$ encircles all poles of the complex plane of $k^{0}$ and $D(k)$ represents the corresponding propagator for the fields  $B_{\mu\nu}$ and $A_{\mu}$. The microcausality condition is violated when there exist poles of $k^{0}$ away from the real axis. 

For the $B_{\mu\nu}$ field, the non-physical pole term Eq. (\ref{antipole}) for the propagator (\ref{prop1}) yields complex poles for $a\neq 0$, and thus violates microcausality condition. However, for the case $a=0$ inspired from the monople solutions \cite{seifert2010a,seifert2010b}, the propagator has no pole contribution from Eq. (\ref{antipole}), and thus the microcausality condition is satisfied. 

In case of vector theory however, a natural consequence of energy positivity violations in the pole terms of propagator (\ref{prop2}), is that there exist poles on the imaginary axis of the complex plane in the $k^0$ integral. The existence of poles away from real axis thus implies violation of microcausality condition. 

\section{\label{sec5}Conclusion}
We derived the dispersion relations for a free antisymmetric tensor field model with spontaneous Lorentz violation as well as an equivalent (at the level of effective actions) vector theory, and checked the conditions for energy positivity violations and microcausality. Depending on the structure of the vacuum value of $b_{\mu\nu}$, the antisymmetric tensor field theory can violate microcausality, while satisfying energy positivity due to the well behaved dispersion relation (\ref{disp2}). For the case wehre $a=0$ in $b_{\mu\nu}$, inspired by the monopole solutions analysed in Refs. \cite{seifert2010a,seifert2010b}, the antisymmetric tensor satisfies the microcausality condition. 

On the contrary, the vector theory (\ref{eq3}) although deemed equivalent to Eq. (\ref{lag}) (at the level of effective action in flat spacetime) yields dispersion relations that violate energy positivity condition, and the propagator possesses poles that cause a violation of the microcausality condition. This indirectly validates the results of Ref. \cite{aashish2019b} showing the breaking of equivalence of vector and antisymmetric theory with spontaneous Lorentz violation beyond the classical level.   
\begin{acknowledgments}
This work was partially funded by DST (Govt. of India), Grant No. SERB/PHY/2017041. The author is thankful to Prof. Sukanta Panda for insightful discussions. 
\end{acknowledgments}





\bibliography{ref,references}

\end{document}